\documentclass[pra, twocolumn]{revtex4}
\pdfoutput=1
\usepackage{amsmath}
\usepackage{amssymb}
\usepackage{mathrsfs}
\usepackage{outlines}
\usepackage{graphicx,verbatim}
\usepackage{enumitem}
\usepackage{hyperref}

\newcommand{\mf}{\mathbf}
\newcommand{\mc}{\mathcal}

\newcommand{\p}{\partial}

\newcommand{\bra}[1]{\left\langle #1 \right|}
\newcommand{\ket}[1]{\left|#1\right\rangle}
\newcommand{\braket}[2]{\left\langle#1 |  #2\right\rangle}
\newcommand{\matel}[3]{\left\langle#1| #2 |  #3\right\rangle}
\newcommand{\avg}[1]{\left\langle #1 \right\rangle}

\usepackage{amsthm}
\usepackage{xcolor}
\theoremstyle{remark}

\theoremstyle{definition}

\theoremstyle{plain}

\begin{document}

\title{ A semiclassical theory of phase-space dynamics of interacting bosons}

\author{R.~Mathew}
\affiliation{Joint Quantum Institute, 
University of Maryland and 
National Institute of Standards and Technology,
College Park, Maryland 20742, USA}
\author{E.~Tiesinga}
\affiliation{Joint Quantum Institute and Joint Center for Quantum Information and Computer Science,
National Institute of Standards and Technology and University of Maryland, Gaithersburg, Maryland 20899, USA}

\begin{abstract}

We study the phase-space representation of dynamics of bosons in the
semiclassical regime where the occupation number of the modes is large.
To this end, we employ the van Vleck-Gutzwiller propagator to obtain
an approximation for the Green's function of the Wigner distribution.
The semiclassical analysis incorporates interference of classical paths
and reduces to the truncated Wigner approximation (TWA) when the interference
is ignored. Furthermore, we identify the Ehrenfest time after which the TWA fails. 
As a case study, we consider a single-mode quantum nonlinear oscillator,
which displays collapse and revival of observables.  We analytically
show that the interference of classical paths leads to revivals, an
effect that is not reproduced by the TWA or a perturbative analysis.

\end{abstract}

\maketitle
\section{Introduction}

The crucial difference between quantum mechanics and a statistical theory
based on classical mechanics is the method of computing the transition
probability between an initial and a final state \cite{feynman_QM}. In
the classical theory, the transition probability is the sum over
probabilities of the paths connecting the two states.  In contrast,
in quantum mechanics, the transition probability is obtained by first
summing the amplitudes of all the connecting paths and then squaring
the sum. This procedure leads to interference, a feature absent in the
classical theory.  An archetypal example of interference is a double-slit
experiment in which a beam of particles after passing through two slits
forms an oscillating intensity pattern on a screen \cite{feynman_QM}.

The aforementioned difference between the theories can be systematically
studied in the semiclassical regime (where the typical action
$\gg \hbar$, the reduced Planck's constant).  
In this regime, a probability amplitude
can be approximated by the contributions from a subset of all connecting 
paths: the classical
paths \cite{schulman_book,morette_definition_1951}.  
(This is the case with the
textbook treatment of the double-slit experiment.)  Crucially, within
this semiclassical approximation, the transition probability retains
interference of paths, albeit classical ones.  The role of classical
trajectories in quantum dynamics was first elucidated by van Vleck
\cite{vleck_correspondence_1928}.  Later, Gutzwiller extended the van
Vleck propagator by including Maslov indices and used it to derive his
trace formula \cite{gutzwiller_periodic_1971}.
The role of classical paths in quantum mechanics 
has been extensively studied; for example, 
in scattering \cite{pechukas_time-dependent_1969}, 
localization \cite{rammer_book,brouwer_anderson_2008}, 
quantum kicked rotor \cite{tian_ehrenfest_2005}, 
level statistics \cite{aleiner_role_1997,muller_periodic-orbit_2005},
quantum work \cite{jarzynski_quantum-classical_2015}, the
Helium atom \cite{wintgen_semiclassical_1992} and quantum transport 
\cite{richter_book,baranger_weak_1993}.

In this paper, we study a semiclassical approximation of the phase-space
dynamics of interacting bosons in the Wigner-Weyl representation. In
this representation, unitary evolution of an initial quantum state
in the Hilbert space is equivalent to evolution of an initial Wigner
distribution in phase space in accordance to the Moyal's equation
\cite{groenewold_principles_1946,moyal_quantum_1949,curtright_concise_2014}.
The reduction of the state space from a high-dimensional Hilbert
space to a lower-dimensional phase space makes the phase-space picture
particularly useful for implementing approximations of quantum dynamics.
An approximation that is usually made is a mean-field approach.  In this
case, the distribution is approximated at all times by a delta function whose location
is determined by the classical Hamilton's equations.  The Gross-Pitaevskii
equation and its discrete versions fall under this category.

An improvement over the mean-field description
is the truncated Wigner approximation (TWA)
\cite{heller_cellular_1991,steel_dynamical_1998,blakie_dynamics_2008},
where the initial distribution is extended and is the Wigner transform of
a quantum state.  The subsequent dynamics of the Wigner distribution is
still classical.  Equivalently, the Moyal's equation is replaced by the
classical Liouville's equation.  In the literature, the TWA is sometimes
called a semiclassical method eventhough it lacks interference effects.
Quantum corrections to the TWA for interacting bosons were studied by
A. Polkovnikov \cite{polkovnikov_quantum_2003,polkovnikov_phase_2010}
using a perturbation theory with the TWA as its zeroth-order approximation.
In particular, a nonlinear oscillator was studied whose quantum
dynamics exhibits collapse and revival of coherences. The perturbative
analysis describes the initial collapse, with increasing accuracy
with the order of the perturbation parameter. It fails to describe revivals
in the system because the analysis still lacks interference of classical paths.

We study semiclassical dynamics of a general Bose system in phase space
that incorporates interference of classical paths and makes comparison
with the TWA transparent. In particular, our analysis identifies the
Ehrenfest time associated with the TWA as the time when interference of
classical paths becomes important.  As a case study, we investigate the
nonlinear oscillator and show that the semiclassical dynamics leads
to revivals.  Recently, others have also applied semiclassical methods
to bosons.  For example, these methods have been applied to
coherent backscattering \cite{engl_coherent_2014} and autocorrelation
functions \cite{tomsovic_post-ehrenfest_2017} in the Bose-Hubbard model.
In addition, the semiclassical Herman-Kluck propagator has been used to 
study boson dynamics \cite{simon_time-dependent_2014,ray_dynamics_2016}.

The remainder of the paper is organized as follows.  First, we define
the phase space of a bosonic system and the Green's function of a Wigner
distribution in Sec.~\ref{sec:phase_space} and Sec.~\ref{sec:G_WF},
respectively.  A semiclassical approximation of this Green's function is
obtained in Sec.~\ref{sec:semi_GF}.  In Sec.~\ref{sec:TWA_SC}, we find
that our semiclassical formalism reduces to the TWA when the interference
terms are ignored. Next, we discuss Ehrenfest times associated with
the TWA and semiclassical approximation in Sec.~\ref{sec:Ehrenfest}.
Subsequently, we apply our formalism to analytically study of a
nonlinear oscillator in Sec.~\ref{sec:nonlinear_osc} and conclude in
Sec.~\ref{sec:conclusions_beyond_twa}.

\section{Phase-space formulation of a bosonic system}
\label{sec:phase_space}

A bosonic system with a finite number of modes can be described in terms
of annihilation and creation operators $\hat a_j$ and $\hat a^\dagger_j$,
respectively, with $j= 1, \dots, d$, where $d$ is the number of modes.
For example, the modes could be the sites of a Bose-Hubbard model or
spin components of a single-mode Bose-Einstein condensate.
The operators satisfy the commutation relations $[a_j, a^\dagger_k] =
\delta_{jk}$, where $\delta_{jk}$ is the Kronecker delta function.
To construct the phase space, we first define the quadrature operators
$\hat x_j = \sqrt{\hbar/ 2}\,(\hat a_j + \hat a^\dagger_j)$ and $\hat
p_j = -i\sqrt{ \hbar/ 2}\,(\hat a_j - \hat a^\dagger_j)$ satisfying
the canonical commutation relations $[\hat x_j, \hat p_k] = i \hbar
\delta_{jk}$.  The eigenstates of $\hat {x}_j$ satisfy $\hat x_j \ket{\mf
x} = x_j \ket{\mf x}$ for all $j \in \{1, \cdots, d \}$, with ``position''
$\mf x = (x_1, x_2, \dots, x_d)$.  Similarly, the eigenstates of $\hat
p_j$ satisfy $\hat p_j \ket{\mf p}= p_j \ket{\mf p}$, with ``momentum''
$\mf p = (p_1, p_2, \dots, p_d)$. 
The eigenstates form a complete basis with $\braket{\mf x'}{\mf x}
= \delta(\mf x - \mf x')$, $\braket{\mf p'}{\mf p} = \delta(\mf p -
\mf p')$ and $\int d\mf x \, \ket{\mf x}\bra{\mf x} = \int d\mf p \,
\ket{\mf p}\bra{\mf p} = 1$, where $\delta(\mf z)$ is a Dirac delta function
and the integrals are over $\mathbb{R}^d$.
We construct a phase space by imposing $\{x_i, p_j \} =  \delta_{ij}$,
where $\{., .\}$ is the Poisson bracket.  We will refer to $\mf r =
(\mf x, \mf p)$ as a phase-space point.  Thus, by introducing quadrature
operators, we have mapped the kinematics of a many-body boson system
with $d$ modes to that of a single particle in $d$-dimensional position
or configuration space.

The Wigner transform \cite{curtright_concise_2014,hillery_distribution_1984} 
maps an operator $\hat
{\mc O}$, a function of $\hat a_j$ and $\hat a^\dagger_j$ or $\hat x_j$
and $\hat p_j$, to its Weyl symbol $\mc O$ in the phase space.  
In fact,
\begin{equation}
    \mc O(\mathbf r) = \int d\mathbf q\,
    \matel{\mf x + \tfrac 1 2\mf q}{\hat{\mc O}}{\mf x - \tfrac 1 2 \mf q}
    e^{-i \mf p \cdot \mf q/\hbar}, \label{eq:Weyl_sym_def}
\end{equation} 
where $\mathbf r = (\mathbf x, \mathbf p)$, ${\mf p \cdot \mf q}$ is the
dot product between $\mf p$ and $\mf q$, and the integral is over the
configuration space $\mf q \in \mathbb{R}^d$.  In particular, the Wigner
distribution $W(\mf r, t)$ at time $t$ is the Weyl symbol of the density
operator $\hat \rho(t)$, up to a factor of $1/(2\pi \hbar)^d$, i.e.,
\begin{equation}
    W(\mathbf r, t) = \frac{1}{(2\pi \hbar)^d} \int d\mathbf q\,
    \matel{\mf x + \tfrac 1 2\mf q}{\hat{\rho}(t)}{\mf x - \tfrac 1 2
    \mf q} e^{-i \mf p \cdot \mf q/\hbar}, \label{eq:defW}
\end{equation}
These definitions imply that $\int d\mf r\, W(\mf r, t) =1$ and 
in the Schr\"odinger picture, the expectation value 
of an operator $\hat {\mc O}$ at a 
time $t$ is
\begin{equation}
    \avg{\hat{\mc O}(t)}\equiv \operatorname{Tr}[\hat \rho(t) \hat{\mc O}] 
    = \int d\mf r\, W(\mf r, t) \mc O(\mf r),
    \label{eq:avgO}
\end{equation}
where the integrals are over the phase space $\mathbb{R}^{2d}$.
Equivalently, in the Heisenberg picture,
\begin{equation}
    \avg{\hat{\mc O}(t)}\equiv \operatorname{Tr}[\hat \rho \, \hat{\mc O}(t)] 
    = \int d\mf r\, W_0(\mf r) \mc O(\mf r, t),
    \label{eq:avgO_H}
\end{equation}
where $W_0(\mf r)$ is the initial Wigner distribution.

\section{Green's function of the Wigner distribution}
\label{sec:G_WF}

The Green's function $G(\mathbf r_f, \mf r_i, t)$ 
of the Wigner distribution in the Schr\"odinger picture is defined by 
\cite{moyal_quantum_1949,berry_quantum_1979,marinov_new_1991,dittrich_semiclassical_2010} 
\begin{equation}
    W(\mathbf r_f, t) = \int d\mathbf r_i\, G(\mathbf r_f, \mf r_i, t) W(\mf r_i, 0),
    \label{eq:defG}
\end{equation}
for $t \geq 0$ with $\mf r_f = (\mf x_f, \mf p_f)$, $\mf r_i = (\mf x_i, \mf p_i)$ 
and 
$G(\mathbf r_f, \mf r_i, 0) = \delta(\mathbf r_f - \mf
r_i)$. In a seminal paper on quantum dynamics in phase space, Moyal called
$G(\mathbf r_f, \mf r_i, t)$ the ``temporal transformation function''
\cite{moyal_quantum_1949}. He derived an expression for $G(\mathbf r_f,
\mf r_i, t)$ in terms of Feynman propagators.  We give a short and direct 
derivation.

The time evolution of the density operator is $\hat \rho(t) = \hat U(t)
\hat \rho_0 \hat U^\dagger(t)$, where $\hat U(t)$ and $\hat\rho_0$ are the unitary
time-evolution and the initial density operator, respectively. We
insert 
$
\int d\mf y_1 \ket{\mf y_1} \bra{\mf y_1}=1
$ and 
$
\int d\mf y_2 \ket{\mf y_2} \bra{\mf y_2}=1
$, with $\mf y_1$ and $\mf y_2$ in configuration space,
into Eq.~\ref{eq:defW} and
find
\begin{align}
    W(\mathbf r_f, t) 
    &= 
    \frac{1}{(2\pi \hbar)^d} \int d\mathbf q d\mathbf y_1 d\mathbf y_2\,
    K\left(\mathbf x_f+ \tfrac 1 2\mathbf q\,, \mathbf y_1, t \right)
    \label{eq:W1}
    \\
    &\quad\times 
    \matel{\mf y_1}{\hat\rho_0}{\mf y_2} 
    K^*\left(\mathbf x_f - \tfrac 1 2\mathbf q, \mathbf y_2, t \right)
    e^{-i \mathbf p_f \cdot \mathbf q/\hbar}\, , \nonumber
\end{align}
where 
$K(\mf x_f, \mf x_i, t) =
\bra{\mf x_f} \hat U(t) \ket{\mf x_i}$
is the Feynman propagator in the configuration space.  For notational
simplicity, we have and will hereafter set $\hbar=1$.  Next, we
express the initial condition $\matel{\mf y_1}{\hat\rho_0}{\mf y_2}$
in terms of the initial Wigner distribution. To this end, we multiply
Eq.~\ref{eq:defW}, evaluated at $t=0$ and $\mf r = \mf r_i$, by $e^{i
\mf p_i \cdot \mf q'}$ and integrate over $\mf p_i$ to find
\begin{equation}
    \int d\mathbf p_i\, e^{i \mathbf p_i \cdot \mf q'/\hbar} W(\mf r_i, 0)
    =
    \matel{\mf x_i + \tfrac 1 2 \mathbf q'}
    {\hat \rho_0}
    {\mf x_i - \tfrac 1 2 \mathbf q'}
\end{equation}
We substitute this expression in Eq.~\ref{eq:W1} and identify $\mf y_1
= \mf x_i +\tfrac 1 2 \mathbf q' $ and $\mf y_2 = \mf x_i -\tfrac 1 2
\mathbf q'$.
From the definition of Green's function 
in Eq.~\ref{eq:defG} we find
\begin{multline}
    G(\mf r_f, \mf r_i , t) = \frac{1}{(2\pi \hbar)^d}\int d\mf q d\mf q' \,
    K\left(\mathbf x_f+ \tfrac 1 2\mathbf q\,, \mathbf x_i 
    + \tfrac 1 2 \mathbf q'\,, t \right) \\
    \times K^*\left(\mathbf x_f - \tfrac 1 2\mathbf q\,, \mathbf x_i - \tfrac 1 2\mathbf q', t \right)
    e^{- i [\mf p_f \cdot\mf q -  \mf p_i \cdot\mf q']/\hbar}.
    \label{eq:G_exact}
\end{multline}
Thus, the exact Green's function of the Wigner distribution involves the
product of two Feynman propagators in configuration space.  We expect
that this product will have interference terms.

\section{Semiclassical approximation of the Green's function}
\label{sec:semi_GF}

A quantum system is said to be in the semiclassical regime when
the typical action (in units of $\hbar$) that appears in the path
integral description of the Feynman propagator is much greater
than one.  For bosonic modes, this regime corresponds to large
occupation numbers.  In fact, the semiclassical approximation of
the propagator, also known as the van Vleck-Gutzwiller propagator,
is \cite{littlejohn_van_1992,schulman_book}
\begin{equation}
    K_{\rm SC}(\mf x_f, \mf x_i, t) 
    = 
    \sum_b \frac{e^{-i \mu^b \pi/2}}{(2\pi i \hbar)^{d/2}}
    \sqrt{\mc D^b(\mf x_f, \mf x_i, t)}
    e^{i S^b(\mf x_f, \mf x_i, t)/\hbar},
    \label{eq:vanvleck}
\end{equation}
where the sum is over all classical paths, indexed by $b$, that start
from position $\mf x_i$ and reach $\mf x_f$ in time $t$.  The action
$
S^b(\mf x_f, \mf x_i, t) 
= \int_0^t d\tau\, L[\mf x_{\rm cl}^b(\tau), d \mf x_{\rm cl}^b(\tau)/d\tau]
$, 
where $L$ is the system Lagrangian and $\mf x^b_{\rm cl}(\tau)$ is
the position as a function of time $\tau$ of the $b$-th classical
path  with $\mf x^b_{\rm cl}(0) = \mf x_i$ and $\mf x^b_{\rm cl}(t) =
\mf x_f$. Finally, $\mu^b$ is the Maslov index and $ \mathcal D^b(\mf x_f,
\mf x_i, t) = |\det\left[ \p^2 S^b(\mf x_f, \mf x_i, t)/ (\p \mf x_f\p
\mf x_i)\right]| $ is the absolute value of the determinant of a $d\times
d$ matrix.

\begin{figure}
  \begin{center}
    \includegraphics[width=0.9\columnwidth]{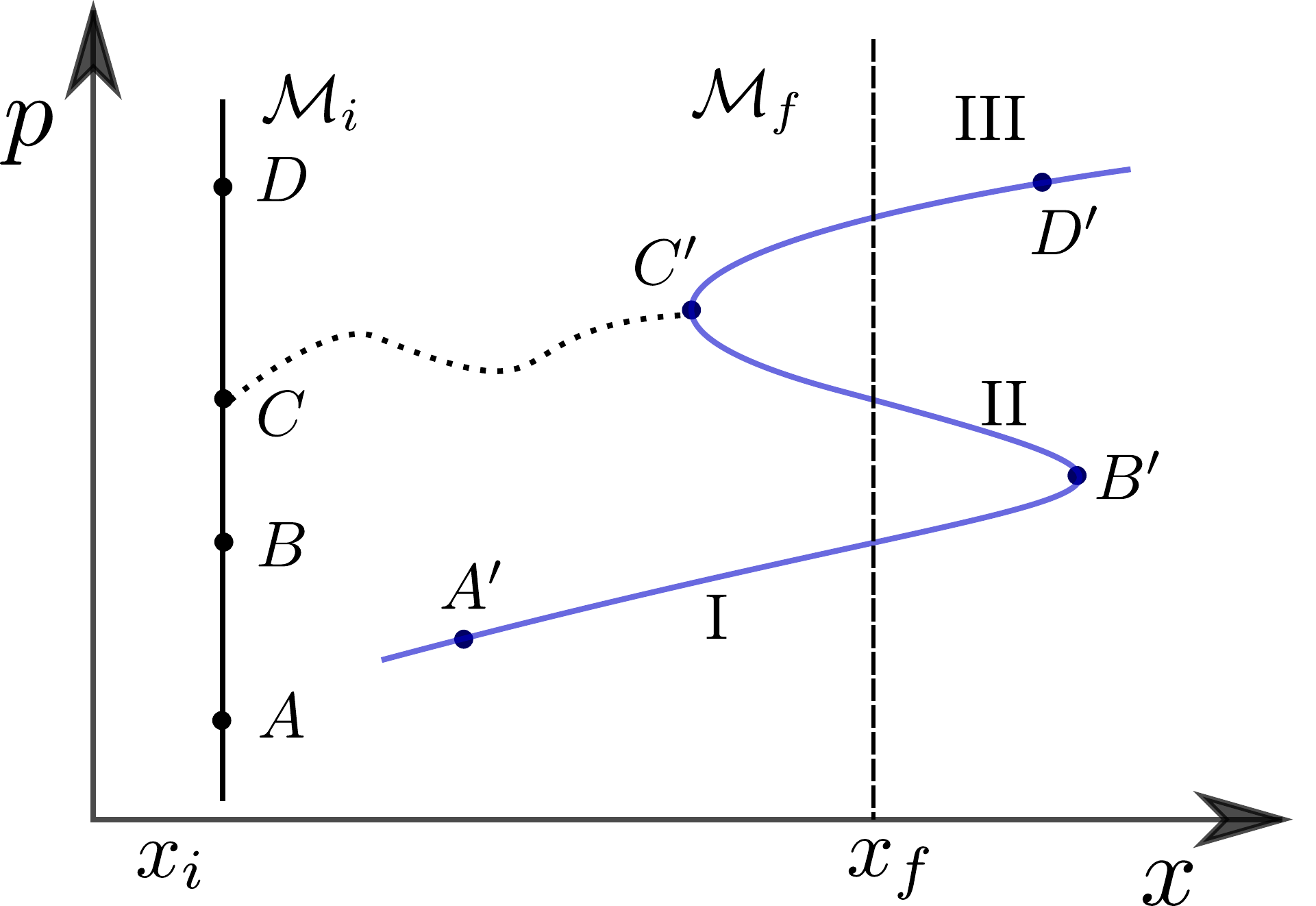}
  \end{center}
    \caption[Classical dynamics of a manifold.]{
Classical dynamics of a manifold in a two-dimensional phase space $(x, p)$.
An initial manifold $\mc M_i$, the line $x=x_i$ shown in black,
evolves into the manifold $\mc M_f$, the curve shown in blue, at time $t$. 
For example, 
points $A$, $B$, $C$, and $D$
on $\mc M_i$ are mapped to $A'$, $B'$, $C'$ and $D'$ on $\mc M_f$, respectively.
The classical path connecting $C$ and $C'$ is also shown.
The manifold $\mc M_f$ has three branches I, II and III separated 
by caustics $B'$ and $C'$.
The line $x=x_f$ intersects 
$\mc M_f$ thrice;  therefore, 
three paths start on $\mc M_i$ and reach position $x_f$ 
at time $t$.
    }
    \label{fig:lag}
\end{figure}

The number of classical paths contributing to $K_{\rm SC}(\mf x_f,
\mf x_i, t)$ can be found by studying the dynamics of the initial
Lagrangian manifold $\mc M_i$, the hyperplane $\mf x= \mf x_i$. Classical
evolution of each point of $\mc M_i$ yields a final manifold $\mc M_f$.
Figure \ref{fig:lag} shows $\mc M_i$ and $\mc M_f$ for a two-dimensional
example.  The final manifold folds at singular positions $\mf x_f$ on $\mc
M_f$ where the number of momenta $\mf p_f$ on $\mc M_f$ as a function of
$\mf x_f$ changes.  Parts of the manifold $\mc M_f$ between these singular
regions are ``branches''.  The example in Fig.~\ref{fig:lag}  has the
three such branches.  
Crucially, the final momentum $\mf p_f(\mf
x_f, \mf x_i, t)$, which, in general, is a multivalued function of $\mf
x_f$ at fixed $\mf x_i$ and $t$, is unique on each branch. Therefore,
classical paths connecting $\mf x_i$ and $\mf x_f$ can be indexed by
the branches that intersect the manifold $\mf x = \mf x_f$.  It is
these branches which contribute to the van Vleck-Gutzwiller propagator
in Eq.~\ref{eq:vanvleck}.  For example, for the position $x_f$ shown in
Fig.~\ref{fig:lag} has three paths that contribute to the propagator.

Substitution of van Vleck-Gutzwiller propagator in Eq.~\ref{eq:G_exact} yields
the semiclassical approximation to the Green's function
\begin{widetext}
\begin{align}
    G_{\rm SC}(\mf r_f, \mf r_i , t) 
    &= 
    \frac{1}{(2\pi \hbar)^{2d}}\int d\mf q d\mf q' \,
    e^{- i [\mf p_f \cdot\mf q -  \mf p_i \cdot\mf q']/\hbar} 
\sum_b 
\sqrt{\mathcal D^b(\mf x_f+ \mf q/2, \mf x_i + \mf q'/2, t)}\,
    e^{i S^b(\mf x_f+ \mf q/2, \mf x_i + \mf q'/2, t)/\hbar-i\mu^b \pi/2}\nonumber\\
&\times \sum_{b'} 
\sqrt{\mathcal D^{b'}(\mf x_f- \mf q/2, \mf x_i - \mf q'/2, t)}\,
    e^{-i S^{b'}(\mf x_f- \mf q/2, \mf x_i - \mf q'/2, t)/\hbar+i\mu^{b'}\pi/2}.
 \label{eq:G_SC}   
\end{align}
\end{widetext}
The expression is cumbersome for our analytical study.  
To proceed, we assume that in Eq.~\ref{eq:G_SC} only the contributions
from small regions $Q$ and $Q'$ around $\mf q =0$ and $\mf q' =0$,
respectively, are important; and, secondly,
the Taylor expansion of the action 
\begin{align}
S^b\left(\mf x_f+ \tfrac 1 2\mf q, \mf x_i + \tfrac 12 \mf q', t\right) 
    \approx S^b(\mf x_f, \mf x_i, t) 
    + \frac{ {\mf p^b_f \cdot\mf q} }{2}
    - \frac{ {\mf p^b_i \cdot\mf q'}}{2},
    \label{}
\end{align}
up to linear terms is sufficient in these regions.
Here, 
$\mf p^{b}_{i} = -\p S^b(\mf x_f, \mf x_i, t)/ \p \mf x_{i}  $ 
and $\mf p^b_f = \p S^b(\mf x_f, \mf x_i, t)/ \p \mf x_f$, respectively, 
are the initial and final momenta 
of the classical path along which the action is computed (see Appendix~\ref{app:D_Sc}
for a derivation). 
We further assume that the extent of the small 
regions $Q$ and $Q'$ in each direction in position space
is much greater than $\sqrt{\hbar}$. 
(Note that from Sec.~\ref{sec:phase_space}, both
position and momentum have the same units as $\sqrt{\hbar}$).
Furthermore, we approximate 
$    
\mathcal D^b\left(\mf x_f \pm \tfrac 12 \mf q, \mf x_i \pm \tfrac 12\mf q', t\right)
$
by
$
\mathcal D^b(\mf x_f, \mf x_i , t)
$.
Substituting these approximations for $S^b$ and $\mc D^b$ in Eq.~\ref{eq:G_SC}
and interchanging the sum and the integral, we find 
\begin{widetext}
\begin{align}
    G(\mf r_f, \mf r_i , t) 
    &= 
    \frac{1}{(2\pi \hbar)^{2d}}
    \sum_{b, b'}
\sqrt{\mathcal D^b\left( \mf x_f, \mf x_i, t \right) 
    \mc D^{b'}\left( \mf x_f, \mf x_i, t \right)}\,
e^{i [S^b\left( \mf x_f, \mf x_i, t \right) - 
S^{b'}\left( \mf x_f, \mf x_i, t \right)]/\hbar- i (\mu^b-\mu^{b'})\pi/2} \nonumber\\
& \quad \quad \times \int_{Q} d\mf q \int_{Q'}d\mf q' \, 
    e^{- i \left( \mf p_f - \mf p^b_f/2 - \mf p^{b'}_f/2\right)\cdot\mf q/\hbar 
     + i \left( \mf p_i - \mf p^b_i/2 - \mf p^{b'}_i/2\right) \cdot\mf q'/\hbar }.
 \label{eq:G_SC2}   
\end{align}
\end{widetext}
Implicit in the existence of $Q$ is the assumption that
$\mf x_f$ is away from the position of any caustics, where two branches
meet. The example in Fig.~\ref{fig:lag} has two caustics.

The integration over $\mf q$ and $\mf q'$ yields
functions of $\mf p_f$ and $\mf p_i$ localized
around $\mf p_i = \tfrac 12 (\mf p_i^b + \mf p_i^{b'})$
and $\mf p_f = \tfrac 12 (\mf p_f^b + \mf p_f^{b'})$,
whose characteristic widths in momentum space are much less than $\sqrt{\hbar}$.
(For ``rectangular'' regions $Q$ and $Q'$
we obtain multidimensional sinc functions.)
Typically, observables are 
smooth functions in phase space, i.e., they vary slowly 
on the scale of $\sqrt{\hbar}$. 
Moreover, initial states of interest 
are classical states (coherent states) whose width is of the order of $\sqrt{\hbar}$.
( We do not consider initial Wigner distributions with fine sub-Planck structures.)
Then we can approximate the localized functions by $\delta$-functions to find
\begin{widetext}
\begin{align}
    G_{\rm SC}(\mf r_f, \mf r_i , t) 
    \approx
    \sum_{b , b'} 
    \sqrt{\mc D^b \mc D^{b'}}
    e^{i (S^b - S^{b'})/\hbar-i(\mu^b - \mu^{b'})\pi/2}\,
    \delta\left[\mf p_f - \tfrac 1 2 (\mf p^b_f + \mf p^{b'}_f) \right]
    \delta\left[\mf p_i - \tfrac 1 2 (\mf p^b_i + \mf p^{b'}_i) \right],
    \label{eq:G_SC3}
\end{align}
\end{widetext}
where, for clarity, we suppress the dependence of $S^b$, $\mc D^b$, etc., 
on $\mf x_i$, $\mf x_f$ and $t$.
This is the main result of this paper and relates the Green's function
of the Wigner distribution to a double sum over classical paths connecting
positions $\mf x_i$ and $\mf x_f$ in time $t$.

\subsection{The truncated Wigner approximation}
\label{sec:TWA_SC}

In the TWA, the Wigner distribution is propagated classically, i.e.,
it obeys the Liouville's equation.  
The Green's function according to the Liouville's equation is 
\begin{equation}
G_{\rm TWA}(\mf r_f, \mf r_i, t) 
=
\delta[\mf x_f - \mf x_{\rm cl}(t; \mf r_i)]
\delta[\mf p_f - \mf p_{\rm cl}(t; \mf r_i)],
\label{eq:GTWA}
\end{equation}
where $[\mf x_{\rm cl}(t; \mf r_i), \mf p_{\rm cl}(t; \mf r_i)]$ 
is the classical path starting from $\mf r_i = (\mf x_i, \mf p_i)$.  
We now show that the ``diagonal'' part of the double sum in
Eq.~\ref{eq:G_SC3}, i.e., when $b= b'$, is equal to $G_{\rm TWA}$.
To this end, we change the independent variables $(\mf x_i, \mf p_i, t)$
of Eq.~\ref{eq:GTWA} to $(\mf x_f$, $\mf x_i, t)$
and find
\begin{equation}
G_{\rm TWA}(\mf r_f, \mf r_i, t)
=
\sum_b  \frac{\delta\left( \mf p_i -  \mf p_i^b \right)}
{\left|\det\left[   \frac{\p \mf x_{\rm cl}(t; \mf r_i)}{\p \mf p_i}
    \bigg\rvert_{\mf p_i = \mf p_i^b} \right]\right|}
\delta(\mf p_f - \mf p_f^b ),
    \label{eq:GTWA2}
\end{equation}
where the sum is over all roots $\mf p_i^b$ (enumerated by $b$) of 
equation 
$\mf x_{\rm cl}(t; \mf x_i, \mf p_i) = \mf x_f$, 
and $\mf p_f^b = \mf p_{\rm cl}(t; \mf x_i, \mf p^b_i)$
\footnote{
The equation is the multidimensional version of the formula
$\delta(z(y)-z_0) = \sum_i \delta(y-y_i)/|z'(y_i)| $, where the sum
is over the roots $y_i$ of the equation $z(y) = z_0$ and $z'(y)$ is the
derivative of $z$ with respect to $y$. 
}.
We have suppressed the dependence of 
$\mf p_i^b$ and $\mf p_f^b$ on $(\mf x_f, \mf x_i, t)$.
Next, we apply the inverse function
theorem, which states that the matrix inverse of a Jacobian is the Jacobian
of the inverse mapping, to find
\begin{align}
    \frac{1} 
    {\det\left[ \frac{\p \mf x_{\rm cl}(t; \mf r_i)}{\p \mf p_i}
    \bigg\rvert_{\mf p_i = \mf p_i^b} \right]}
    = 
    \det\left[ -  \frac{\p^2 S^b(\mf x_f, \mf x_i, t)}
        {\p \mf x_f \p\mf x_i}
    \bigg\rvert_{\mf x_f = \mf x_f^b} \right],
\end{align}
where we used that $\mf p_i^b = -\p S^b(\mf x_f, \mf x_i, t)/ \p \mf x_{i}$.
Substituting the expression in Eq.~\ref{eq:GTWA2},
we arrive at 
\begin{equation}
    G_{\rm TWA}(\mf r_f, \mf r_i ,t) 
    = 
    \sum_b \mc D^b(\mf x_f, \mf x_i,
    t) \, \delta\left( \mf p_f - \mf p^b_f  \right) \delta\left( \mf p_i -
    \mf p^b_i  \right),
\end{equation} 
which is the diagonal part of Eq.~\ref{eq:G_SC3}.
Thus, TWA ignores interference of classical paths.
For the special cases of the harmonic oscillator and free particle,
the TWA matches with the quantum motion because only a single path contributes
to the sum in Eq.~\ref{eq:G_SC3} and, hence, there are no interference terms.

\subsection{Ehrenfest times}

\label{sec:Ehrenfest}

An Ehrenfest time is the time scale when an approximation to
the quantum motion deviates appreciably from exact evolution
\cite{berman_condition_1978,chirikov_quantum_1988}.  In fact, there
is a hierarchy of Ehrenfest times based on the approximations to the
quantum dynamics \cite{silvestrov_ehrenfest_2002,tomsovic_comment_2003}.
For the mean-field approximation, the Ehrenfest time $\tau_{\rm MF}$  is
the time scale when an initially localized Wigner distribution becomes
distorted and stretched due to nonlinear (not necessarily chaotic)
classical dynamics.  From Sec.~\ref{sec:TWA_SC}, we find that the
Ehrenfest time $\tau_{\rm TWA}$ associated with the TWA occurs when
interference of classical paths becomes important. This time scale
is greater than $\tau_{\rm MF}$ because interference of paths occurs
when the Wigner distribution becomes so distorted that it fills up the
accessible phase space. Finally, there is $\tau_{\rm SC}$, the Ehrenfest
time for the breakdown of semiclassical approximation based on van
Vleck-Gutzwiller propagator, which is greater than $\tau_{\rm TWA}$.
Numerical studies have shown that the breakdown occurs when diffraction
becomes important \cite{tomsovic_long-time_1993,dittes_long-time_1994}.

\section{Case study: A nonlinear oscillator}
\label{sec:nonlinear_osc}

We consider a single-mode nonlinear oscillator whose quantum Hamiltonian is
\begin{equation}
    \hat H_{\rm NO} =  \frac{U}{2}\hat a^\dagger \hat a^\dagger \hat a \hat a,
    \label{eq:HNO}
\end{equation}
where $U$ is the interaction strength and $\hat a(\hat a^\dagger)$ is
the annihilation (creation) operator of the associated bosonic mode.
As the number operator $\hat a^\dagger \hat a$ commutes with $H_{\rm
NO}$, the energy eigenstates are $\ket n$ with eigen-energies $E_n = U
n(n-1)/2$, where $n$ is the occupation number of the mode.  Decomposing an
arbitrary initial state $\ket{\psi_0} = \sum_{n=0}^\infty c_n \ket n$
and noting that $n(n-1)/2$ is an integer, we can immediately see that
the time-evolved state  $\ket{\psi(t)}$ periodically revives, i.e.,
$\ket{\psi(t)}= \ket{\psi_0}$ when $t$ is an integer multiple of the
period $t_{\rm rev} = 2\pi \hbar/U$. 

\begin{figure}
  \begin{center}
    \includegraphics[width=0.9\columnwidth]{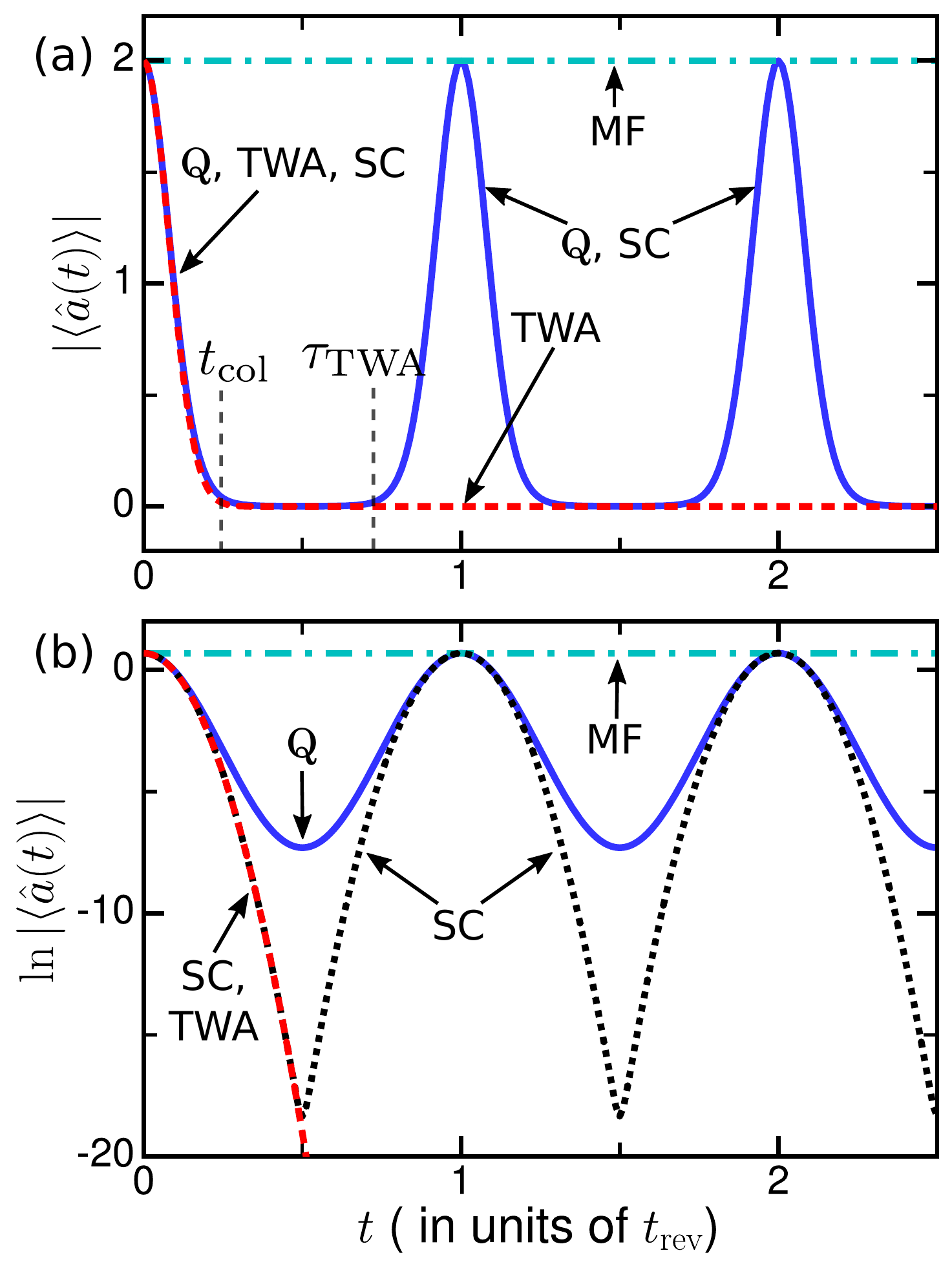}
  \end{center}
    \caption[Collapse and revival in a nonlinear oscillator]{
        Collapse and revival in a nonlinear oscillator.  Panels (a) and
        (b) show $|\avg{a(t)}|$ and $\ln|\avg{a(t)}|$, 
        respectively, as a function of time $t$ for an initial
        coherent state whose mean atom number is four. Exact quantum
        dynamics (labeled Q) displays  collapse and revival of
        $|\avg{a(t)}|$ with collapse and revival times $t_{\rm col}$ 
        and $t_{\rm rev}$, respectively. 
        The mean-field 
        solution, labeled MF, is time independent.
        The TWA result,
        Eq.~\ref{eq:at_TWA}, closely replicates the first collapse
        but shows no revival and deviates appreciably from the quantum dynamics 
        after a time $\tau_{\rm TWA}$.
        On the other hand, the semiclassical
        approximation (labeled SC), Eq.~\ref{eq:at_SC}, agrees well
        with the quantum evolution for all times. In panel (a)
        the semiclassical and quantum curves are indistinguishable. 
    }
    \label{fig:collapse_and_revival}
\end{figure}

The nonlinear oscillator has been studied in experiments with a BEC in
an optical lattice \cite{greiner_collapse_2002} and with 
photons using Kerr nonlinearity \cite{kirchmair_observation_2013}.
In these experiments,
the initial state is well-described by a 
coherent state, $\ket{\psi_0}= e^{-|\alpha|^2/2} \sum_{n=0}^\infty \alpha^n/\sqrt{n!}
\ket n$, where $\alpha$, in general, is a complex number
and $|\alpha|^2 = N$
is the average number of atoms or photons. 
Using interference, 
the collapse and revival of 
the absolute value of the 
expectation value of $\hat a$ and  a generalized Husimi function, respectively, 
were measured in these experiments.
We find that
the expectation value of $\hat a$ evolves as
\begin{align}
    \avg{\hat a(t)} = \matel{\psi(t)}{\hat a}{\psi(t)} 
         &= 
         \alpha e^{|\alpha|^2 (e^{- i  U t/\hbar} - 1)}.
    \label{eq:at_exact}
\end{align}
Its absolute value is shown in
Fig.~\ref{fig:collapse_and_revival}. 
At short times $U t/\hbar \ll 1$, 
\begin{equation}
\avg{\hat a(t)}
\approx  \alpha e^{-  |\alpha|^2 U^2 t^2/(2 \hbar^2) - i U|\alpha|^2 t/\hbar},
    \label{eq:at_collapse}
\end{equation}
whose decay in time is Gaussian with time constant
$\hbar/(U\sqrt{N})$. 
The collapse time $t_{\rm col}$ is a few times this time constant,
as shown in Fig.~\ref{fig:collapse_and_revival}, and is much smaller
$t_{\rm rev}$ for large $N$.  In the experiments with a BEC in an optical
lattice, three-body effects proportional to $(\hat a^\dagger)^3 \hat a^3$
change the nature of the collapse and revival in an interesting manner
\cite{johnson_effective_2009,will_time-resolved_2010,tiesinga_collapse_2011}.

\subsection{Dynamics according to the TWA}
\label{sec:TWA_NO}

Next, we study the time dynamics of $\avg{\hat a(t)}$
within the TWA. First, we need to write down
the classical Hamiltonian corresponding to $\hat H_{\rm NO}$
\footnote{
We use $\hat a^\dagger \hat a = (\hat x^2 + \hat p^2)/(2\hbar) - 1/2$ 
to find
that 
$
\hat H_{\rm NO} 
= 
U/8 \times[(\hat x^2 + \hat p^2)^2/\hbar^2 - 4(\hat x^2 +
\hat p^2)/\hbar + 3]
$. 
We then replace $\hat x$, $\hat
p$ by their classical limits to obtain $\mc H_{\rm NO}$, and ignore the
second and third terms in the semiclassical limit $N \gg 1$.
}. It is
\begin{equation}
    \mc H_{\rm NO} =  \frac{U}{8 \hbar^2} (x^2 + p^2)^2
    \label{eq:Hcl_NO}
\end{equation}
with classical equations of motion 
\begin{equation}
    \frac{dx}{dt} = \frac{\p \mc H_{\rm NO} }{\p p} = \frac{U}{2 \hbar^2} \rho^2 p \,
    , \quad
    \frac{dp}{dt} = -\frac{\p \mc H_{\rm NO} }{\p x} 
    = - \frac{U}{2 \hbar^2} \rho^2 x,
    \label{}
\end{equation}
where $\rho^2 = x^2+p^2$. Hence,
classical paths lie on  circles
in phase space centered around the origin $(x, p) = (0, 0)$. 
Their oscillation frequencies are $\omega = U \rho^2/(2 \hbar^2)$.  
The system is
integrable as the phase space is two-dimensional and energy
is conserved.  The angle of the action-angle coordinates is the polar
angle $\varphi$ measured in clockwise direction of
motion and evolves as $\varphi(t) = \omega t + \varphi_i$, 
where $\varphi_i$ is the initial angle.
Using the definition $\omega = \p \mc H_{\rm NO}/\p I$,
we find that the action coordinate $I=\rho^2/2$. Hence, 
$\rho$ is a constant of motion.

For concreteness, let the initial coherent state $\ket{\psi_0}$,
with occupation number $N \gg 1$, be centered along
the $p$-axis,  i.e., $\alpha=i\sqrt{N}$.
Its Wigner distribution
\begin{equation}
    W_0(x, p) = \frac{1}{\pi \hbar}e^{- \left[ x^2 + (p-\sqrt{2 \hbar N})^2 \right]/\hbar},
    \label{eq:W0_NO}
\end{equation}
is centered at $(x, p) = (0, \sqrt{2\hbar N })$ and width $O(\sqrt{\hbar})$.
Next, we calculate $\avg{\hat a(t)}_{\rm TWA}$, 
the expectation value of $\hat a$ within the TWA.
Instead of using the Green's function $G_{\rm TWA}$ of Eq.~\ref{eq:GTWA}, 
it 
is more convenient to work in the Heisenberg picture. In this picture, 
the Wigner-Weyl transform of the operator $\hat a(t)$ 
is $a(t) =
[x(t) + ip(t)]/\sqrt{2 \hbar} = i \rho e^{-i\varphi(t)}/\sqrt{2 \hbar}$
with $\varphi \in (-\pi, \pi]$ and $\varphi = 0$ 
along the $p$-axis.  
Then using Eq.~\ref{eq:avgO_H} and writing $W_0(x, p)$ in polar coordinates,
we find 
\begin{align}
    \avg{\hat{a}(t)}_{\rm TWA} 
    &=  \int_0^\infty \rho d\rho 
    \int_{-\pi}^\pi  d\varphi_i \, \frac{i\rho}{\sqrt{2\hbar}} 
        e^{-i \varphi(t)} W_0(\rho, \varphi_i).
    \label{eq:at_TWA}
\end{align}
For $N \gg 1$, it is sufficient to expand the exponent of $W_0(x, p)$ 
to second order in $\rho$ and $\varphi$ around the location of the maximum of
the Wigner distribution, i.e.,
\begin{equation}
W_0(\rho, \varphi_i) 
\approx 
\frac{1}{\pi \hbar}e^{- \left[(\rho- \sqrt{2 \hbar N})^2 - 2\hbar N\varphi_i^2 \right]/\hbar}.
    \label{eq:W0_NO_polar}
\end{equation}
Substituting this expression and $\omega = U \rho^2/(2\hbar)$  in
Eq.~\ref{eq:at_TWA}, we derive
\begin{align}
    \avg{\hat{a}(t)}_{\rm TWA} 
    &\approx i \sqrt{N} 
     e^{- U^2 N t^2/(2 \hbar) - i U N t/\hbar},
    \label{eq:at_TWA2}
\end{align}
which matches the initial collapse of the coherent state in
Eq.~\ref{eq:at_collapse}, but has no revival. A comparison
of Eq.~\ref{eq:at_TWA2} with the exact quantum result of
Eq.~\ref{eq:at_exact} for the absolute value of $\avg{\hat a(t)}$ is
shown in Fig.~\ref{fig:collapse_and_revival}.  The figure also shows the
mean-field value $|\avg{\hat a(t)}_{\rm MF}|$, 
which is $|a(t)|$ along the single circular trajectory starting from 
$(x, p) = (0, \sqrt{2\hbar N})$. 
 Thus, $|\avg{\hat a(t)}_{\rm MF}| = \sqrt{N}$ is a constant.

The classical phenomenon of phase-space mixing explains the collapse of
$\avg{\hat a(t)}$ \cite{few_mode_paper}.  For an integrable system, the
coarse-grained long-time Wigner distribution is uniformly distributed in
the angle coordinates of the action-angle variables.  For the nonlinear
oscillator, $a(t) \propto e^{i \varphi(t)}$ and its expectation value
goes to zero as the Wigner distribution mixes in the angle $\varphi$.
Furthermore, within the TWA, the coarsened Wigner distribution reaches
a steady state; hence, there is no revival. This latter observation
indicates that quantum interference reverses phase-space mixing and
revives the quantum state. In the next section, we find that applying
the semiclassical formalism, indeed, leads to revival.

\subsection{Dynamics according to the semiclassical approximation}
\label{sec:at_SC}

The calculation of $\avg{\hat a(t)}_{\rm SC}$ according to the
semiclassical approximation is lengthy and has been relegated to
Appendices~\ref{app:S_nonlinear_osc} and \ref{app:at_SC}. 
We first calculate the action in terms of the polar angles and
winding number of classical paths around the origin in Appendix
\ref{app:S_nonlinear_osc}.  We carry out the remainder the calculation
in Appendix~\ref{app:at_SC}. Here, we list the main steps:
\begin{enumerate}
\item
The time evolution of an observable with Weyl symbol 
$\mc O(x, p)$ in the Schr\"odinger picture is given by Eq.~\ref{eq:avgO}.
We replace $G(r_f, r_i, t)$ by 
$G_{\rm SC}(r_f, r_i, t)$ as given in Eq.~\ref{eq:G_SC3} and carry out 
the integrals over $p_i$ and $p_f$ to arrive at Eq.~\ref{eq:Ot_SC1}.

\item 
The classical equations of motion are 
simplest in the action-angle
coordinates. Therefore, 
we convert the integrals over $x_i$, $x_f$ and the double sum 
over $b$, $b'$ in Eq.~\ref{eq:Ot_SC1} 
into integrals over the initial and
final angles $\varphi_i$ and $\varphi_f$,
respectively, and a double sum over winding numbers of classical
paths around the origin.  
We also express the observable, $G_{\rm SC}(r_f,  r_i ,t)$, and the initial
Wigner distribution in terms of $\varphi_i$, $\varphi_f$ and winding numbers.

\item 

Next, we note that the classical motion in the phase space is restricted
in an annulus of radius $\sqrt{2N}$ and width of $O(1)$.  Then,
$\mc O(x, p) \approx \mc
O(\sqrt{2N}\sin\varphi, \sqrt{2N} \cos\varphi)$; in particular, $a(x,
p) \approx i\sqrt{N} e^{-i\varphi}$.  We make similar approximations
for the determinants $\mc D^b(x_f, x_i, t)$. 
The initial Wigner distribution, however, varies sharply with $\rho$ and 
requires a more careful approximation.
We then solve the remaining integrals.
\end{enumerate}

Finally, we find
\begin{equation}
\avg{\hat a(t)}_{\rm SC}
=
\sum_{v=-\infty}^\infty i\sqrt{N} e^{-i UNt/\hbar } 
    e^{- \left(2\pi v - U t/\hbar\right)^2 N/2 },
\label{eq:at_SC}
\end{equation}
where $v$ is difference of the winding number of the interfering 
paths.
This expression
corresponds to a train of localized Gaussians and 
is invariant under the transformation $t \to t + 2\pi \hbar/U$
and $v \to v -1$; hence, is periodic with time period $t_{\rm rev}
= 2\pi \hbar/U$.
Figure \ref{fig:collapse_and_revival} shows that
$\avg{\hat a(t)}_{\rm SC}$ agrees with the  
exact quantum average $\matel{\psi(t)}{\hat a}{\psi(t)}$ 
for all times.

Finally, we discuss the Ehrenfest times of the nonlinear oscillator for
initial coherent states.  From Fig.~\ref{fig:collapse_and_revival}(a), we
see that the mean-field prediction deviates from the quantum evolution
well before the collapse time $t_{\rm col}$, i.e., $\tau_{\rm MF}
< t_{\rm col}$.  In contrast, the deviation of the TWA from quantum
evolution (ignoring exponentially small differences) occurs abruptly
after a finite time $\tau_{\rm TWA} = t_{\rm rev} - t_{\rm col}$ before
the first revival of $\avg{\hat a(t)}$.  The interference of classical
paths starts at $\tau_{\rm inter}$ when the Wigner distribution fills
up the annular accessible phase space, i.e.,  when paths starting from
the initial localized distribution with winding numbers zero and one
terminate in the same small region of phase space.  We can estimate
$\tau_{\rm inter}$ by noting that for a coherent state the distribution
of classical frequencies $\omega$ has a mean $ U N/\hbar$ and width
$\Delta \omega = U \sqrt{N}/\hbar$.  Therefore, $\tau_{\rm inter} \sim
2\pi/\Delta \omega$ is of the order of $t_{\rm col}$ and, hence, is much
smaller than $\tau_{\rm TWA}$.  In other words, it takes time for the
interference of paths to affect $\avg{\hat a(t)}$ appreciably. In fact,
at $\tau_{\rm TWA}$ the number of interfering classical paths is of the
order of $\sqrt{N}$. On the other hand, the Ehrenfest time $\tau_{\rm SC}$ is infinite
for the nonlinear oscillator.

The Ehrenfest time $\tau_{\rm TWA}$ depends on the observable under
consideration.  For example, for $\avg{\hat a^2(t)}$ the collapse and
revival times are $t_{\rm rev}/2$ and $t_{\rm col}/2$, respectively,
and the TWA fails after $(t_{\rm rev} - t_{\rm col})/2$.  Nevertheless,
$\tau_{\rm TWA}$ is still greater than $\tau_{\rm inter}$ for all
observables (that are polynomials in $a$ and $a^\dagger$ with a degree
smaller than $N$).  We also expect the delay in effects of interference
and the dependence of $\tau_{\rm TWA}$ on the observable to hold true for
generic integrable systems (where the dynamics is away from singularities
like a saddle point of the classical Hamiltonian).  In contrast, in
a chaotic system and for motion near a saddle point of an integrable
system, the Ehrenfest time $\tau_{\rm TWA} \sim \tau_{\rm inter}$
\cite{aleiner_divergence_1996,rozenbaum_lyapunov_2017,few_mode_paper}.

\section{Conclusion and outlook}
\label{sec:conclusions_beyond_twa}

In conclusion, we presented a semiclassical theory of phase-space
dynamics of bosons.  We derived a semiclassical approximation,
Eq.~\ref{eq:G_SC}, to the exact Green's function of the Wigner
distribution.  Crucially, the approximation preserves
the quantum interference of classical trajectories. In fact, we have
shown that the formalism reduces to the TWA when the interference terms
are ignored.  Hence, the Ehrenfest time associated with the breakdown of
the TWA occurs when interference of classical paths becomes important.
As a case study, we examined a single-mode nonlinear oscillator whose
exact quantum dynamics exhibits collapse and revival.  We investigated
the dynamics of an observable of this oscillator using the TWA and
our semiclassical formalism.  Within TWA, the expectation value of an
observable collapses due to phase mixing, and there is no revival. The
semiclassical approximation, however, reproduces revivals and accurately
matches the exact quantum dynamics for all times.

Finally, we comment on the long-time validity of our semiclassical
approximation.  For the nonlinear oscillator, the
semiclassical approach is valid for all times
\footnote{The time evolution of observables that are polynomial in
$a$ and $a^\dagger$ can be obtained by a generalization of the analysis
in the appendix.}.  We expect this to be true for generic integrable
systems as they can be quantized by the Einstein-Brillouin-Keller
method \cite{keller_corrected_1958}.  The situation, however, is
not straightforward for chaotic systems.  
For example,
the semiclassical evolution (based on the
van Vleck-Gutzwiller propagator) of an initial wavefunction defined on a Lagrangian
manifold, whose Wigner distribution is not localized,
breaks down after a time of the order of the
Ehrenfest time associated with interference of classical paths
\cite{berry_evolution_1979,berry_quantum_1979}. For localized
initial Wigner distributions, however, numerical studies and heuristic
arguments have shown that the van Vleck-Gutzwiller propagator
works  for rather longer times
\cite{tomsovic_long-time_1993,dittes_long-time_1994,heller_postmodern_1993}
and only breaks down due to diffraction.  The validity of
our semiclassical approach for chaotic systems will require further study.

\appendix

\section{Derivatives of action} 
\label{app:D_Sc}

We evaluate the partial derivatives of the action $S^b(\mf x_f, \mf
x_i, t)$ with respect to the initial and final positions.  The action
satisfies the Hamilton-Jacobi equation and, in principle, its derivates are well
known \cite{goldstein_book,arnold_book}.  Here, we 
give a derivation for
the sake of completeness.  For notational simplicity, we assume that
the configuration space is one-dimensional; generalization to higher
dimensions is straightforward.  Consider a classical path $[x^b_{\rm
cl}(\tau), p^b_{\rm cl}(\tau)]$, which starts from the phase-space point
$(x_i, p^b_i)$ and ends at $(x_f, p^b_f)$. Next,
consider another classical path whose position in time,
$x_{\rm cl}^b(\tau) + \delta x_{\rm cl}^b(\tau)$, 
is infinitesimally close to $x_{\rm cl}^b(\tau)$
such that $\delta x_{\rm cl}^b(0) = \Delta x_i$ and $\delta x_{\rm cl}^b(t) = 0$.  
Then the
change in the action is
\begin{align*}
    \Delta S^b  
    &= 
    \int_0^t d\tau\, \left( \frac{\p L}{\p x_{\rm cl}} \delta x_{\rm cl}^b(\tau) 
    + \frac{\p L}{\p \dot x_{\rm cl}} \delta \dot x_{\rm cl}^b(\tau) \right)\\
    &=
    \frac{\p L}{\p \dot x_{\rm cl}} \delta x_{\rm cl}^b(\tau)\bigg\rvert_0^t
    +
    \int_0^t dt\, \left( \frac{\p L}{\p x_{\rm cl}} 
    - \frac{d}{dt}\frac{\p L}{\p \dot x_{\rm cl}} \right)\delta x_{\rm cl}^b(\tau),
    \label{}
\end{align*}
where $\dot x_{\rm cl} = dx_{\rm cl}/d\tau$ and we have suppressed the arguments of
$L$.  Now, the second term vanishes because $x_{\rm cl}^b(\tau)$ satisfies the
Euler-Lagrange equations of motion.  Using the fact that 
$p = \p L(x, \dot x)/\p \dot x$, we have $\Delta S^b = -p^b_i \Delta x_i$ or
\begin{equation}
    \frac{\p S^b(x_f, x_i, t)}{\p x_i} = -p^b_i.
    \label{eq:deriv_Sc}
\end{equation}
Similarly, we can prove that
\begin{equation}
    \frac{\p S^b(x_f, x_i, t)}{\p x_f} = p^b_f.
    \label{}
\end{equation}

\section{Action of the nonlinear oscillator} 

\label{app:S_nonlinear_osc}

\begin{figure}
    \centering
    \includegraphics[width=0.9\columnwidth]{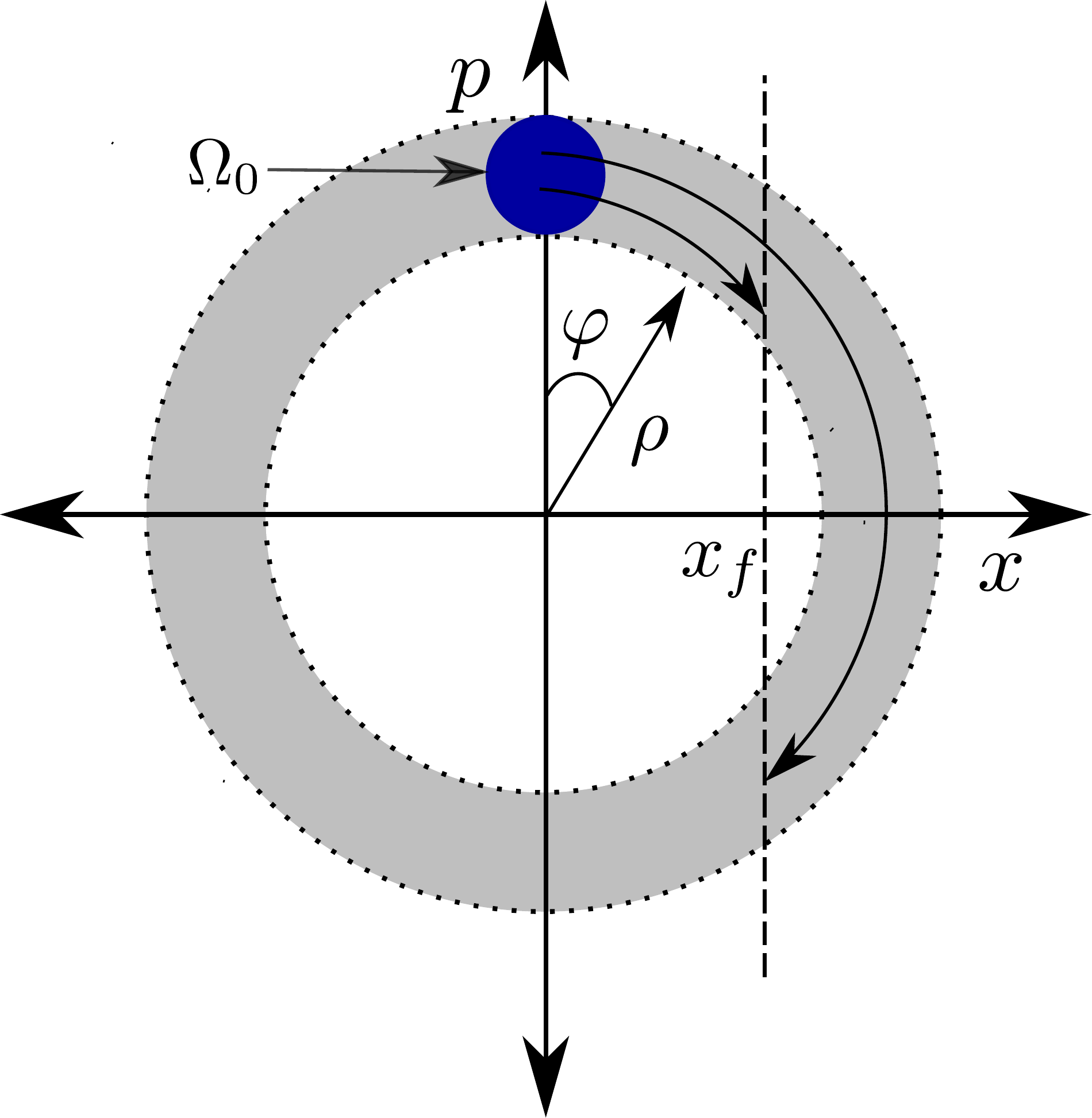}
    \caption{
Classical paths of a nonlinear oscillator in phase space $(x, p)$
starting from a Wigner distribution initially 
localized in region $\Omega_0$ shown by the solid blue circle.  
Polar coordinates $(\rho, \varphi)$, with
angle $\varphi$ measured from the $p$-axis in a clockwise direction,
are also shown.
The region $\Omega_0$ is centered at $(\rho, \varphi) = (\sqrt{2N}, 0)$
and has a width of $O(1)$.
Trajectories starting
from $\Omega_0$ lie within the gray annulus.  Two paths
with traversal time $t$ and winding number zero that start from $x=0$
and end at $x=x_f$ are shown. 
    }
    \label{fig:phase_space_NO}
\end{figure}

We compute the action $S^b(x_f, x_i, t)$ of the nonlinear oscillator
described in Sec.~\ref{sec:nonlinear_osc}.  The action depends on the
index $b$, which we have not yet quantified.  
A natural guess is the
winding number $w$ of a circular path around the phase-space origin. 
The winding number is a nonnegative integer as the motion in phase space is 
unidirectional.
For a given
$(x_f, x_i, w, t)$, however, more than one classical path can exist. 
For example, two such
paths are shown in Fig.~\ref{fig:phase_space_NO}. In contrast,
a given $(\varphi_f, \varphi_i, w, t)$, where $\varphi_i$
and $\varphi_f$ are the initial and final angles, respectively,
uniquely determines a classical path.  
The reason is 
that the 
oscillation frequency is specified by
\begin{equation}
\omega = \frac{(\varphi_f - \varphi_i)\bmod 2\pi + 2\pi w}{t}
    \label{eq:omega_w_rel}
\end{equation}
and, hence, uniquely determines the radius $\rho = \hbar
\sqrt{2\omega/U}$ (see Sec.~\ref{sec:TWA_NO}) of the classical path.

It is convenient to define
the action 
$\mathscr S^w(\varphi_f, \varphi_i, t) = S^b[x_f(\varphi_f,
\varphi_i, w, t) , x_i(\varphi_f, \varphi_i, w, t), t]$, 
indexed by the winding number $w$ of path $b$
and
\begin{align}
    \mathscr S^w(\varphi_f, \varphi_i, t) 
    &= \int_0^t d\tau\, \left[  p_{\rm cl} \frac{dx_{\rm cl}}{d\tau} - \mc H_{\rm NO} \right],
\end{align}
where $\mc H_{\rm NO}$ is given by Eq.~\ref{eq:Hcl_NO} and we have suppressed 
the arguments $(\varphi_f, \varphi_i, w, t)$.
Substituting $x_{\rm cl}(\tau) = \rho \sin \varphi(\tau)$ 
and $p_{\rm cl}(\tau) = \rho \cos\varphi(\tau)$, we find
\begin{align}
    \mathscr S^w(\varphi_f, \varphi_i, t) 
    &=  \int_0^t d\tau\, 
    \left(  \omega\rho^2  \cos^2\varphi(\tau) - \frac{U\rho^4}{8}
    \right)\nonumber,
\end{align}
where we used $d\rho/d\tau = 0$, $d\varphi/d\tau= \omega$ and have 
set $\hbar= 1$.
The integration over $\tau$ yields
\begin{eqnarray}
    \lefteqn{ \mathscr S^w(\varphi_f, \varphi_i, t) 
    =  
    \frac{[(\varphi_f-\varphi_i)\bmod{2\pi} + 2\pi w]}{2Ut}}
    \label{eq:action_NO}
    \\
   && \times
    \left[  
    (\varphi_f-\varphi_i)\bmod{2\pi} + 2\pi w
     + \sin(2\varphi_f) - \sin (2\varphi_i )
    \right].\nonumber
\end{eqnarray}


\section{Calculation of $\avg{\hat a(t)}$}
\label{app:at_SC}

We calculate the expectation value of $\hat a(t)$ within the
semiclassical approximation and follow the outline 
presented in Sec.~\ref{sec:at_SC}.

\begin{enumerate}[wide, labelwidth=!, labelindent=0pt]
\item
The semiclassical evolution of the expectation value of an observable
of the nonlinear oscillator 
$\hat{\mc O}$  with Weyl symbol $\mc O(x, p)$ is 
\begin{align}
    \avg{\hat{\mc O}(t)}_{\rm SC} 
    & = 
    \int dr_i dr_f\, \mc O(r_f) G_{\rm SC}(r_f, r_i, t)
    W_0(r_i).
\end{align}
Substituting $G_{\rm SC}(r_f, r_i, t)$ from Eq.~\ref{eq:G_SC}
and integrating over the momenta $p_i$ and $p_f$, we find
\begin{eqnarray}
    \lefteqn{\avg{\hat{\mc O}(t)}_{\rm SC} 
    = 
    \int dx_i dx_f\, 
    \sum_{b, b'}
    \mc O\left( x_f, \frac{p^b_f + p^{b'}_f}2 \right) }
    \label{eq:Ot_SC1}
    \\
    && \times W_0\left( x_i, \frac{p^b_i + p^{b'}_i}2 \right)
    \sqrt{\mc D^b\mc D^{b'}}\,
    e^{i S^b -i S^{b'} - i(\mu^b - \mu^{b'})\pi/2}, \nonumber
\end{eqnarray}
where we suppress the dependence of $p_i^b$, $\mc D^b$, $S^b$,
etc., on $(x_f, x_i, t)$ and set $\hbar = 1$. The range of integration 
is $(-\infty, \infty)$ for both $x_i$ and $x_f$.

\item
The action has a simpler form in terms of the angles
(see Eq.~\ref{eq:action_NO}). Hence, we proceed to change the integration
variables in Eq.~\ref{eq:Ot_SC1} to the angle coordinates.  To this end,
we first introduce a 
set of initial and final positions $x'_i$ and $x'_f$, respectively, and write
a symmetric expression 
\begin{widetext}
\begin{align}
    \label{eq:Ot_SC2}
    \avg{\mc O(t)}_{\rm SC} 
    &= 
    \int dx_i dx_f dx'_i dx'_f\,  
    \delta(x_i - x'_i) 
    \delta(x_f - x'_f)\\ 
    &\times\sum_{b, b'}
    \mc O\left[ \frac{x_f + x'_f}2, \frac{p^b_f(x_f, x_i, t) + p^{b'}_f(x'_f, x'_i, t)}2 \right] 
    W_0\left[ \frac{x_i + x'_i}2, \frac{p^b_i(x_f, x_i, t) + p^{b'}_i(x'_f, x'_i, t)}2 \right]
    \nonumber
    \\
    &\times
    \sqrt{\mc D^b(x_f, x_i, t) \mc D^{b'}(x'_f, x'_i, t)}\,
    e^{i S^b(x_f, x_i, t) -i S^{b'}(x'_f, x'_i, t) - 
    i[\mu^b(x_f, x_i, t) - \mu^{b'}(x'_f, x'_i, t)]\pi/2} \nonumber,
\end{align}
\end{widetext}
where the explicit dependence of the quantities is shown to avoid any confusion.
The two sets of paths indexed by $b$ and $b'$ now have different boundary 
conditions $(x_f, x_i, t)$ and $(x'_f, x'_i, t)$, respectively,
enabling us to interchange the sum over $b$ and integrals over $x'_i$ 
and $x'_f$.
The next step is to change the integration measure in terms of one for 
the angles.
This step is carried out in Appendix ~\ref{app:x_to_phi} and we find
\begin{align}
    \int dx_i dx_f \sum_b (\dots)
    &= 
    \sum_{w=w_{\rm min}}^{w_{\rm max}} 
    \int_{-\pi}^\pi d\varphi_f \int_{-\pi}^\pi d\varphi_i \, |\det J|\,
    (\dots),
    \label{eq:dx_to_dphi}
\end{align}
where $(\dots)$ is a function of $(x_f, x_i, b, t)$ and the Jacobian matrix
$
J = 
\p (x^w_i, x^w_f)/\p (\varphi_i, \varphi_f)
$
with 
$x^w_i = x_i(\varphi_f, \varphi_i, w, t)$ and 
$x^w_f = x_f(\varphi_f, \varphi_i, w, t)$.
The nonnegative integers $w_{\rm min}$ and $w_{\rm max}$ are minimum
and maximum winding numbers, respectively, of trajectories starting
from region $\Omega_0$, as shown in Fig.~\ref{fig:phase_space_NO}.
An equation analogous to Eq.~\ref{eq:dx_to_dphi} holds for measures
of $x'_i$ and $x'_f$.  Substitution of these measure changes in
Eq.~\ref{eq:Ot_SC2} yields
\begin{widetext}
\begin{align}
    \avg{\mc O(t)}_{\rm SC}
    &=
    \sum_{w, w' = w_{\rm min}}^{w_{\rm max}}
    \int 
    d\varphi_i d\varphi_f
    d\varphi'_i d\varphi'_f\,
    \left|\det\left[\frac {\p (x^w_i, x^w_f)} 
            {\p (\varphi_i, \varphi_f)}\right]\right|
    \left|\det\left[\frac {\p (x^{w'}_i, x^{w'}_f)} 
            {\p (\varphi'_i, \varphi'_f)}\right]\right|
    \delta(x_i^w - x_i^{w'})
    \delta(x_f^w - x_f^{w'})
    \nonumber \\
    &\quad \times
    \mc O\left( \frac{x^w_f + x^{w'}_f}2, \frac{p^w_f + p^{w'}_f}2\right)
    W_0\left(\frac{x^w_i + x^{w'}_i}2, \frac{p^w_i + p^{w'}_i}2 \right)
    \sqrt{\mathscr D^w \mathscr D^{w'}}\,
    e^{i \mathscr S^w -i \mathscr S^{w'} - i(\mu^w - \mu^{w'})\pi/2},
    \label{eq:Ot_SC3}
\end{align}
\end{widetext}
where the arguments of quantities with superscript $w$ and $w'$ are
$(\varphi_i$, $\varphi_f, t)$ and $(\varphi'_i$, $\varphi'_f, t)$,
respectively.  Moreover, we have introduced $\mathscr D^w(\varphi_f, \varphi_i,
t) = \mc D^b(x^w_f, x^w_i, t)$ and $\mathscr S^w(\varphi_f, \varphi_i,
t)$ is given by Eq.~\ref{eq:action_NO}.

\item

We explicitly write all quantities
appearing in Eq.~\ref{eq:Ot_SC3} in terms of $(\varphi_f, \varphi_i, w, t)$.
We do so by noting that
the relevant classical motion is
restricted in an annulus of width $O(1)$ around $\rho = \sqrt{2N}$ 
(see Fig.~\ref{fig:phase_space_NO}).  
In the annulus, we approximate the radius by its mean $\sqrt{2N}$,
i.e., 
$x^w_i \approx \sqrt{2N} \sin \varphi_i$, $p^w_i \approx
\sqrt{2N} \sin \varphi_i$, etc., which leads to 
\begin{equation}
    \left|\det\left[\frac{\p (x^w_i, x^w_f)} 
        {\p (\varphi_i, \varphi_f)}\right]\right|
    \approx 2N |\cos\varphi_i \cos\varphi_f|
\end{equation}
and
\begin{equation}
\mathscr D^w(\varphi_f, \varphi_i, t) 
\approx 
\frac 1 {2 U N t |\cos\varphi_i\cos\varphi_f|},
\end{equation}
etc.
Moreover,
as the initial Wigner
distribution is localized around angle $\varphi=0$,
$
    \delta(x_i^w - x_i^{w'}) 
    \approx 
    \delta(\varphi_f - \varphi_f')/
(\sqrt{2N}\cos\varphi_f).
$
The other delta function becomes
\begin{align}
    \delta(x_f^w - x_f^{w'}) 
    &\approx 
    \frac{
    \delta(\varphi_f - \varphi_f') +
    \delta(\varphi_f+\varphi_f' - \pi) 
}
{\sqrt{2N}\cos\varphi_f}.
\label{eq:delta_phif}
\end{align}
The two contributions
reflect the fact that a line at fixed value of $x_f$
intersects the thin annulus in two regions, whose respective angles 
are approximated by the angle of the intersection with
the circle of radius $\rho = \sqrt{2N}$.

Substituting these approximations into Eq.~\ref{eq:Ot_SC3} and integrating 
over $\varphi'_i$ and $\varphi'_f$, we find
\begin{align}
    \label{eq:Ot_SC4}
    \avg{\mc O(t)}_{\rm SC}
    =  
    \frac{1}{Ut}
    \sum_{w, w' = w_{\rm min}}^{w_{\rm max}}
    \int d\varphi_i d\varphi_f  \, 
    \mc O\, W_0 \\
     \times \,
     e^{i \mathscr S^w -i \mathscr S^{w'} - i(\mu^w - \mu^{w'})\pi/2},
     \nonumber
\end{align}
where we suppress the arguments of $\mc O$ and $W_0$, and 
neglect the contribution from the second term in
Eq.~\ref{eq:delta_phif}. 
This term leads to
a highly oscillating integrand whose integral is small.
The arguments of quantities in the integrand 
with either superscript $w$ or $w'$
are now $\varphi_f$, $\varphi_i$ and $t$.

Next, we note that $\mc O(x,p)$ is a slowly varying function of
$x, p$ and  within the annulus
$
\mc O\left[
\tfrac 1 2 \left( x^w_f + x^{w'}_f \right) , 
\tfrac 1 2 \left( p^w_f + p^{w'}_f \right)
\right] 
\approx \mc
O(\sqrt{2N} \sin\varphi_f, \sqrt{2N} \cos\varphi_f)
    \label{}
$.
In particular,
\begin{equation}
a\left[ \tfrac 1 2 \left( x^w_f + x^{w'}_f \right), 
    \tfrac 1 2 \left( p^w_f + p^{w'}_f \right) \right]
\approx 
i \sqrt{N}e^{-i\varphi_f}.
\label{eq:at_approx}
\end{equation}
We {\it cannot} make a similar approximation for the initial Wigner distribution, 
i.e., replace $\rho^w$ and $\rho^{w'}$ by $\sqrt{2N}$,
because the distribution varies sharply around $\rho=\sqrt{2N}$.
Instead, we write
\begin{eqnarray}
    \lefteqn{\frac{\rho^w(\varphi_f, \varphi_i, t)}{\sqrt{2N}}    
     = 
    \left[ \frac{(\varphi_f - \varphi_i)\bmod 2\pi 
    + 2 \pi w}{U Nt} \right]^{1/2}} \nonumber\\
    &\approx& 1 + \frac{1}{2}\left[ \frac{(\varphi_f - \varphi_i)\bmod 2\pi
    + 2 \pi w}{U Nt} -1 \right],
\end{eqnarray}
where we used the
relation $\rho =\sqrt{2\omega/U}$ (see Sec.~\ref{sec:TWA_NO}), 
Eq.~\ref{eq:omega_w_rel} and 
performed a Taylor expansion around $\rho^w/\sqrt{2N} = 1$.
We substitute $\rho$ in the initial Wigner distribution of 
Eq.~\ref{eq:W0_NO_polar}
by the Taylor approximation for $(\rho^w + \rho^{w'})/2$,
to find
\begin{eqnarray}
    \lefteqn{
        W_0\left(\frac{x^w_i + x^{w'}_i}2, \frac{p^w_i + p^{w'}_i}2 \right)
    \approx
    \label{eq:W0_approx}}
    \\
    &&
    \frac{1}{\pi}e^{
    -
    \left[(\varphi_f - \varphi_i)\bmod 2\pi + \pi(w+w') - UNt \right]^2
    / {2U^2 N t^2}
    -2N\varphi_i^2} \, . \nonumber
\end{eqnarray}

Also, from Eq.~\ref{eq:action_NO}, we have
\begin{align}
    \mathscr S^w - \mathscr S^{w'} 
    &= 
\frac{ 2\pi}{Ut}
(w-w')[(\varphi_f - \varphi_i)\bmod 2\pi + \pi(w + w')]
     \label{eq:action_diff}
\end{align}
Finally, the Maslov index, which is the number of turning points 
of a classical path, increases by two for every winding.
Therefore, 
\begin{equation}
\mu^w - \mu^{w'} = 2(w -w').
    \label{eq:maslov_diff}
\end{equation}

After substituting $\mc O(x, p) = a(x, p)$, Eqs.~\ref{eq:at_approx},
\ref{eq:W0_approx}, \ref{eq:action_diff}, and \ref{eq:maslov_diff} in
Eq.~\ref{eq:Ot_SC4}, we find
\begin{widetext}
\begin{align}
    \avg{a(t)}_{\rm SC}
    &=  
    \frac{i\sqrt{N}}{\pi Ut}
    \sum_{w, w' = w_{\rm min}}^{w_{\rm max}}
    \int_{-\pi}^{\pi} d\varphi_i\int_{-\pi}^{\pi} d\varphi_f  \, 
    e^{-i\varphi_f -i(w - w')\pi}\,
    e^{
    -
    \left[(\varphi_f - \varphi_i)\bmod 2\pi + \pi(w+w') - UNt \right]^2
    /
    (2U^2 N t^2)
    }\nonumber
    \\
    & \quad \times e^{ -2N\varphi_i^2}
    e^{i 2\pi(w-w')[(\varphi_f - \varphi_i)\bmod 2\pi 
    + \pi(w + w')]/ (U t) }.
\end{align}
\end{widetext}
Next, we extend the limits on $w$ and $w'$ to $[0, \infty)$ and
write the sums over $w$ and $w'$ in terms of $u = w + w'$ and $v = w-
w'$. We combine the sum over $u$ and the integral over $\varphi_f$ 
by defining
$y = (\varphi_f - \varphi_i)\bmod 2\pi + \pi u$, whose 
range is $[0, \infty)$.
We
realize that $e^{-i \varphi_f - i(w-w')\pi} = e^{-i (y +\varphi_i)}$ and
the integrand is separable in $\varphi_i$ and $y$. 
After evaluating the integrals, we arrive at
\begin{align}
    \avg{a(t)}_{\rm SC}
    &=
    \sum_{v=-\infty}^\infty 
    i\sqrt{N} 
    e^{-i UNt }
    e^{-\left(2\pi v - U t\right)^2 N/2 } e^{-1/(8N)} \nonumber
\end{align}
which becomes Eq.~\ref{eq:at_SC} of the main text for large $N$.
\end{enumerate}

\subsection{Derivation of Eq.~\ref{eq:dx_to_dphi}}
\label{app:x_to_phi}

Here, we derive Eq.~\ref{eq:dx_to_dphi}.  We restrict our attention
to paths that start from the phase-space region $\Omega_0$,
in which the initial Wigner distribution is concentrated.
Figure~\ref{fig:phase_space_NO} shows the region $\Omega_0$  for
the nonlinear oscillator. The paths starting within
$\Omega_0$ lie on the annulus shown in the figure.
Now, the winding number of a circular path at
a fixed traversal time is a stepwise increasing function of the
radius.  Let the (time-dependent) winding
numbers of paths that lie on the inner and outer circles of the annulus
be $w_{\rm min}$ and $w_{\rm max}$, respectively, with $w_{\rm min} \leq
w_{\rm max}$.  For a given winding number, there can be two 
paths that start from $\Omega_0$ with position $x_i$ and reach 
position $x_f$ in time $t$.  Figure~\ref{fig:phase_space_NO} shows a pair
of such paths with winding number zero and $x_i=0$.  Moreover,
the paths end in the upper ($p>0$) and lower
($p\leq 0$) halves of the phase space.  Therefore, we can interchange the
integrals over boundary conditions and sum over paths to find
\begin{multline}
    \int dx_i dx_f \sum_b \left(\dots \right)
    = 
    \sum_{\substack{w=w_{\rm min}, {\, \rm upper}} }^{w_{\rm max}}
    \int dx_i dx_f (\dots)\\
    + 
    \sum_{\substack{w=w_{\rm min}, {\, \rm lower}} }^{w_{\rm max}}
    \int dx_i dx_f (\dots),
    \label{eq:dx_idx_f}
\end{multline}
where the labels ``upper'' and ``lower'' indicate paths that
end in the corresponding half of phase space.

In each half of the phase space, the final angle is uniquely determined
given $(x_f, x_i, w, t)$.  Therefore, we can transform the integrals
over $x_i$ and $x_f$ in Eq.~\ref{eq:dx_idx_f}  to one over angles
and combine the ``upper'' and ``lower'' contributions to arrive at
Eq.~\ref{eq:dx_to_dphi}.

\begin{center}
\hfill
$\square$
\end{center}

\bibliography{sc_boson}
\end{document}